**Title:** Experimental Demonstration of Non-Resonant Hyperlens in the Visible Range

**Author Information**

Authors: Jingbo Sun, Mikhail I. Shalaev, and Natalia M. Litchinitser*

Affiliations:

1 Department of Electrical Engineering, University at Buffalo, The State University of New York, Buffalo, NY 14260, USA.

Correspondence and requests for materials should be addressed to natashal@buffalo.edu

**Abstract**: A metamaterial hyperlens offers a unique solution to overcome the diffraction limit by transforming evanescent waves responsible for imaging subwavelength features of an object into propagating waves. However, the first realizations of optical hyperlenses were limited by a narrow working bandwidth and significant resonance-induced loss. Here, we report the first experimental demonstration of a non-resonant waveguide-coupled hyperlens operating in the visible wavelength range. A detailed investigation of various materials systems proves that a radial fan-shaped configuration is superior to the concentric layer-based configuration in that it relies on non-resonant negative dielectric response, and, as a result, enables broadband and low-loss performance in the visible range.

**Main Text:** Since the resolution of conventional optical systems is limited by diffraction, the visualization of features smaller than the wavelength of the illuminating light requires the development of new imaging techniques. Such techniques would revolutionize many fields ranging from clinical diagnostics and single molecule spectroscopy to nanoscale lithography. For instance, it was shown that an ability to visualize nanoscale structures might be critical for early detection of various cancers, such as ovarian cancer, which is the fifth leading cause of death due to cancer in women[1], or adenocarcinoma in patients with chronic gastroesophageal reflux disease[2].

Metamaterials-based optics was predicted to solve the problem of diffraction-limited resolution of conventional optical components[3-10]. One of the most promising approaches to high-resolution applications is a so-called hyperlens[11-14]. Hyperlenses overcome the diffraction limit by transforming evanescent waves responsible for imaging subwavelength features of an object into propagating waves. Once converted, those formerly decaying (evanescent) components commonly lost in conventional optical imaging can now be collected and transmitted using standard optical components[11, 12]. A hyperlens is a curved (cylindrically or spherically shaped) hyperbolic metamaterial with negative dielectric permittivity along the radial direction and a positive permittivity along the tangential direction. To date, the most commonly used design to realize such a hyperbolic metamaterial is the concentric ring multilayered structure[13-18]. The Drude dispersion of metals is always negative below the plasma frequency, which is usually in the deep ultra-violet range for metals, such as silver or gold. Therefore, the only way to achieve the right indefinite property is to obtain negative permittivity by using resonant negative dielectric permittivity in the direction perpendicular to the metal layer and positive dielectric permittivity along the metal layers by choosing the working frequency close to the plasma frequency[19-22]. However, the use of resonant structures results in a limited bandwidth and high losses near the resonance. An alternative approach would rely on having negative dielectric permittivity along metal layers (or metal wires) and positive dielectric permittivity in the orthogonal direction[23-28]. As discussed in detail below, this alternative approach has strong potential to enable broadband, low-loss hyperlenses. However, until now, it had only been designed and demonstrated in the microwave frequency range and acoustics[29~31], while the challenge of fabricating such structures in the optical frequency range (where such a device is of paramount importance) has still not been overcome.

Here, we propose and experimentally demonstrate a non-resonant waveguide-integrated hyperlens with a radially oriented layered structure in the visible frequency range. We performed a detailed investigation of various materials systems and proved that a radial (fan-shaped) configuration is superior to the concentric-layers-based configuration. Indeed, the radial configuration with a negative component of dielectric permittivity orientated along the layers and positive permittivity being normal to the layers relies on non-resonant negative dielectric response and results in a broadband, low-loss performance in the visible range. In such a device, the evanescent-wave components of a wave vector from the subwavelength slits are converted into the propagating waves and then out-coupled to the free space in the far field.

The dielectric permittivity of a hyperlens is described by an indefinite tensor:

$$\tilde{\varepsilon} = \begin{bmatrix} \varepsilon_\rho & 0 \\ 0 & \varepsilon_\theta \end{bmatrix}, \quad \varepsilon_\rho < 0, \quad \varepsilon_\theta > 0, \quad (1)$$

where $\varepsilon_\rho$ is the permittivity component in the radial direction, which in our case corresponds to the direction along the layers, and $\varepsilon_\theta$ is the azimuthal component of the dielectric permittivity. Therefore, the equi-frequency contour (EFC) of such a structure in the $\rho$-$\theta$ plane can be described by

$$\frac{k_\rho^2}{\varepsilon_\theta} + \frac{k_\theta^2}{\varepsilon_\rho} = \frac{\omega^2}{c^2}. \quad (2)$$

In Eq. 2, $k_\rho$, and $k_\theta$ are the wavevector components along the radial and azimuthal directions, respectively; $c$ is the speed of light, and $\omega$ is the frequency of the incident wave.

As shown in Fig. 1, the indefinite tensor in Eq. 1 can be realized in different frequency ranges using metal/dielectric multilayered metamaterial in two orientations:

i) $\varepsilon_\perp = \varepsilon_{r,\theta} > 0, \quad \varepsilon_{//} = \varepsilon_{r,\rho} < 0$, or ii) $\varepsilon_{//} = \varepsilon_\theta > 0, \quad \varepsilon_\perp = \varepsilon_\rho < 0$,

where $\varepsilon_{//}$ is the dielectric permittivity along the layers and $\varepsilon_\perp$ is the dielectric permittivity perpendicular to the layers.

Using the Maxwell-Garnett theory, dielectric permittivity tensor components of the multilayered structure can be determined by

$$\varepsilon_{\parallel} = \varepsilon_d f + \varepsilon_m (1-f), \qquad (3a)$$

$$\varepsilon_{\perp} = \frac{\varepsilon_d \varepsilon_m}{\varepsilon_m f + \varepsilon_d (1-f)}, \qquad (3b)$$

where $f$ is the filling ratio of the dielectric material (e.g., Poly(methyl methacrylate), or PMMA), $\varepsilon_{\parallel}$ is complex dielectric permittivity along the layers, and $\varepsilon_{\perp}$ is complex dielectric permittivity perpendicular to the layers. The frequency dependence of the permittivity of metal can be described by the Drude model as $\varepsilon_m = \varepsilon_{\infty} - \omega_p^2 / [\omega(\omega + i\gamma)]$, while the permittivity of the PMMA, $\varepsilon_d$, is nearly frequency independent. According to Eq. 3a, frequency dependence of dielectric permittivity along the layers, $\varepsilon_{\parallel}$, is dominated by the properties of metal and is described by the Drude-like model with an effective plasma frequency $\omega_{pe}$ (corresponding to the wavelength $\lambda_{pe}$) for the given filling ratio $f$. This effective Drude model shows that for $\lambda < \lambda_{pe}$, $\varepsilon_{r,\parallel} > 0$ and for $\lambda > \lambda_{pe}$, $\varepsilon_{r,\parallel} < 0$. Here, the subscript "$r$" denotes the real part of dielectric permittivity. Moreover, the value of $\lambda_{pe}$ increases with the increasing $f$ of dielectric material. According to Eq. 3b, frequency dependence of dielectric permittivity perpendicular to the layers is described by the Lorentz-like model, and the real part of dielectric permittivity becomes negative ($\varepsilon_{r,\perp} < 0$) at wavelengths below the resonance.

Here, we investigate indefinite dielectric properties of three multilayered structures consisting of one of three metallic materials (gold[32], silver[32], and TiN[33]), and a dielectric material ($\varepsilon_d$=2.1, e.g., PMMA or MgF$_2$) in different wavelength ranges and with different layer orientations. The dielectric permittivity components (real part) of such multilayered structures for different filling ratios can be calculated using Eq. 3 and are shown in Fig. 2. Figure 2a, c and e, and Fig. 2b, d and f, show $\varepsilon_{r,\parallel}$ and $\varepsilon_{r,\perp}$ as functions of wavelength and filling ratio for the three combinations, respectively. The dashed black curves indicate the $\lambda_{pe}(f)$ for $\varepsilon_{/\!/}$ in these three cases. Negative $\varepsilon_{r,\perp}$ regions resulting from the Lorentz resonance are enclosed by the dotted black curves.

Figure 2b shows that the Au/dielectric multilayer possesses indefinite properties only in one orientation: $\varepsilon_{r,/\!/}<0$, $\varepsilon_{r,\perp}>0$, which is in the Region "a" between the dashed and dotted black curves, and it covers the entire visible range provided that a proper filling ratio is chosen. In contrast, the indefinite property of the Ag/Dielectric and TiN/Dielectric can be obtained in two orientations.

As shown in Fig. 2d and f, there are two regions ("a" and "b") located in between the dashed and dotted black curves. In the case of Ag/dielectric multilayer, Region a, corresponding to $\varepsilon_{r,/\!/}<0$, $\varepsilon_{r,\perp}>0$, completely covers the visible range, while Region b, corresponding to $\varepsilon_{r,/\!/}>0$, $\varepsilon_{r,\perp}<0$, only appears in the ultra-violet (UV) frequency range. Similar to the Ag/dielectric case, in the case of TiN/dielectric, both orientations are possible, but $\lambda_{pe}$ is significantly larger. In particular, Region a, corresponding to $\varepsilon_{r,/\!/}<0$, $\varepsilon_{r,\perp}>0$ for the TiN/dielectric multilayer, covers the range of $\lambda>640$nm. Region b, corresponding to $\varepsilon_{r,/\!/}>0$, $\varepsilon_{r,\perp}<0$, is rather narrow and only extends from 510nm to 640nm. These results are summarized in Table 1.

From the fabrication view, a multilayered structure with $\varepsilon_{r,/\!/}>0$, $\varepsilon_{r,\perp}<0$ (Region b) is easier to realize. However, as it is shown in Table 1, this orientation corresponds to a limited bandwidth and high losses owing to the resonant nature of negative permittivity response (see the also imaginary part of the permittivity dispersions in Supplementary Materials). For the Ag/dielectric case, Region b only covers the UV frequency range. For the TiN/dielectric multilayer, Region b is broader; however, the material losses are significantly larger than those of Ag (in addition to the losses associated with the resonance). On the other hand, based on the data in Fig. 2 and Table 1, a multilayer with $\varepsilon_{r,//}<0$, $\varepsilon_{r,\perp}>0$ (Region a) possesses significantly wider bandwidth (extending over the entire visible range in the cases of Ag or Au) and significantly lower losses owing to the non-resonant origin of the negative dielectric permittivity component $\varepsilon_{r,//}$. These conclusions motivated us to take the challenge of designing and realizing a hyperlens based on $\varepsilon_{r,//}<0$, $\varepsilon_{r,\perp}>0$ orientation. Here, we demonstrate a non-resonant hyperlens with a radially oriented layered structure in the visible frequency range using a combined top-down and bottom-up fabrication approach.

From the above discussion, we chose Au/PMMA materials combination with a PMMA filling ratio of 60%. In this case, the non-resonant indefinite properties ($\varepsilon_{r,\rho}=\varepsilon_{r,//}<0$, $\varepsilon_{r,\theta}=\varepsilon_{r,\perp}>0$) are expected to be observed in the range from 500 to 1000nm. The schematic and actual images of the hyperlens that we designed and fabricated in this work are shown in Fig. 3. The hyperlens was composed of 35 pairs of Au/PMMA layers arranged in a fan-like shape and integrated with an MIM waveguide on a chip, as shown in Fig. 3a. The entire structure, including the hyperlens itself and a blocking layer with two nano-slits, was first made in a 400nm thick PMMA layer placed on top of a gold film using the standard electron-beam lithography (EBL) technique (including e-beam exposure, development, and lift-off steps). As a result, fan-shaped PMMA walls with air slots between them were made as show in Fig. 3b. Next, gold was filled into these air slots using an electroplating method. Figure 3c shows the cross section of the hyperlens structure after electroplating. On top of the PMMA, a 300nm thick layer of silver was deposited in order to form the MIM consisting of the top silver layers, the PMMA guiding layer, and the bottom gold film. As a result, the hyperlens was directly integrated with the MIM waveguide. Finally, a Focused Ion Beam (FIB), was used to mill the input and output ports, consisting of a grating coupler and two arc-shaped slots, respectively (details can be found in the Supplementary Materials).

In order to characterize the performance of the waveguide-coupled hyperlens, the transverse-electric (TE) mode was coupled into the MIM waveguide using the grating coupler on the top silver layer. This mode impinged on the two nano-slits that served as the subwavelength objects to be imaged and split the original single beam into two beams. Due to the hyperbolic dispersion property of the hyperlens, evanescent parts responsible for the subwavelength imaging were preserved (along with propagating waves), converted into propagating waves, and finally, magnified by the hyperlens. Consequently, after the transmission through the hyperlens, waves from the two nano-slits were separated enough to be imaged with conventional optics once they were out-coupled through the output arc-shaped ports.

Taking into account the validity limits of the effective medium approximation and necessary magnification, the size of the hyperlens (outer/inner radii ratio $R_{out}/R_{in}$) was adjusted to a wavelength range of interest. In particular, the size of each Au/PMMA pair should be much smaller than the wavelength of light on both input and output sides of the hyperlens. In this work, we optimized our samples to be used with a 780nm source. The inner and outer radii are 800nm and 2400nm, respectively, enabling 3-times magnification. Two nano-slits with widths of 100nm separated by 300nm were made in the Au blocking layer and were used as the sub-wavelengh objects to be imaged. In our design, the width of one Au/PMMA pair at the outer (output) side was about 200nm, which is still much less than the operating wavelength. At the inner (input) side, each nanoslit size corresponds to almost 3 pairs of Au/PMMA layers. These considerations ensure that the designed multilayered structure can be approximated as the effective medium. According to Eq. 3, the components of the dielectric permittivity for our design are $\varepsilon_\rho = -8.93+0.87i$, $\varepsilon_\theta = 3.65+0.0178i$, corresponding to very low loss. Numerical simulation results based for the structure with these parameters are supplied in the Supplementary Materials section.

In order to characterize the performance of the waveguide-integrated radial hyperlens, we built a twin imaging system shown in Fig. 4a. The location of the input grating coupler (Fig. 4b) was first determined by Imaging System 1 on the right side. Next, a 780nm laser beam was coupled into the waveguide using the same objective lens of the imaging system to focus it on the grating coupler. The output beams were detected by Imaging System 2 on the opposite side of the sample, ensuring that all incident light coming from the top side was blocked. Using this setup, we made certain that the power collected by the second imaging system corresponded to light that passed through the waveguide-hyperlens system.

Figure 4c shows the images taken by Imaging System 2. Two output beams are coupled out of the two output slits corresponding to the two beams from the two nano-slits made in the gold blocking layer and then resolved by the hyperlens. Finally, a reference sample with two nano-slits on the curved gold blocking layer but without a hyperlens was also fabricated and placed inside the MIM waveguide, as shown in Fig. 4e. Figure 4e shows the image taken by Imaging System 2 without the hyperlens clearly showing only one broad beam coming from the middle parts of two arc-shaped slits, which proves that two beams from the nano-slits cannot be resolved without the hyperlens.

In summary, we proposed a non-resonant magnifying hyperlens operating in the visible frequency range and experimentally demonstrated its sub-diffraction limited imaging at 780nm. Non-resonant indefinite dielectric properties enabling low-loss subwavelength imaging at visible wavelengths were realized using a radially oriented metal/dielectric nanoscale layers that were made using a combination of electron beam lithography and electro-plating. By optimizing the outer/inner radii ratio and the filling ratio of the dielectric and metal, the hyperlens can be optimized to work in the entire visible range.

**Acknowledgments:** This research was supported by the US Army Research Office Award # W911NF-11-1-0333 and the U.S. National Science Foundation under the award 1231852.

**Author contributions**

J.S. and N. M. L. proposed the idea developed in this work. J. S. made the design and did the simulation. J. S. and M. S. did the fabrication of the sample. J. S. did the optical characterization of the sample. N. M. L and J. S. wrote the paper. N. M. L. supervised this work.

**Competing financial interests**

The authors declare no competing financial interests.

**Figure legends**

Figure 1 Radial (left) and concentric ring based (right) hyperlens realized using multilayered structures of different orientations.

Figure 2 Real part of parallel ($\varepsilon_{//}$) and perpendicular ($\varepsilon_{\perp}$) components of effective dielectric permittivity as a function of filling ratio ($f$) and wavelength ($\lambda$) for three materials systems: gold/dielectric (top row), silver/dielectric (middle row), and TiN/dielectric (bottom row). The dashed black curves indicate the $\lambda_{pe}(f)$ for $\varepsilon_{r,//}$ in these three cases. Negative $\varepsilon_{r,\perp}$ regions resulting from the Lorentz resonance are enclosed by the dotted black curves.

Figure 3 Optical fan-shaped hyperlens with radial layered structure. (a) Schematic of the hyperlens integrated with a metal-insulator-metal waveguide. The inset shows numerical results of imaging of two sub-diffraction-limited slits. (b) PMMA pattern including the hyperlens and the blocking layer after the EBL. (c) Cross section view of the hyperlens structure after the electroplating of gold. This image is taken at the position indicated by the dashed line in Fig. 3b. The black part is the PMMA, and the grey part is gold, which is filled into the slots in Fig. 3b.

Figure 4 Experimental setup and results: (a) Twin imaging system. (b) SEM image of the entire sample with hyperlens. On the left is the grating coupler used as the input port, and on the right are the output ports of two arc-shaped slits. The hyperlens is shown in between. The capsule shaped protuberant ridge is the excess gold, which has expanded out of the PMMA structure during the electroplating. (c) Direct visualization at the output port taken using the CCD camera. (d) SEM image of the sample in the absence of the hyperlens. (e) Results without the hyperlens.

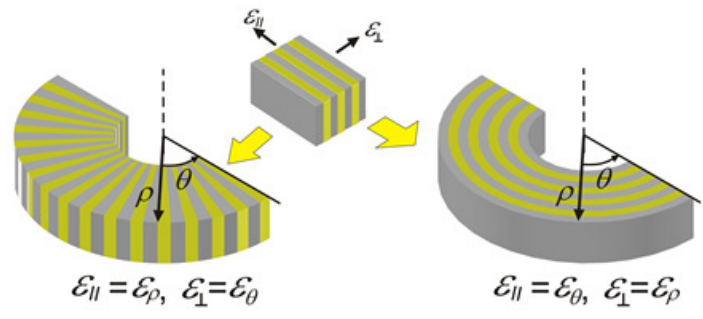

**Figure 1**

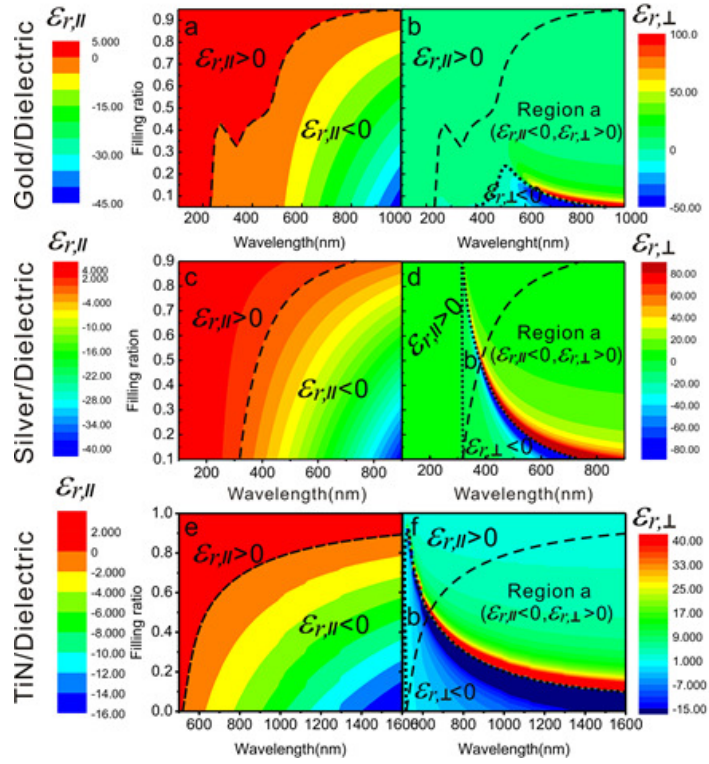

**Figure 2**

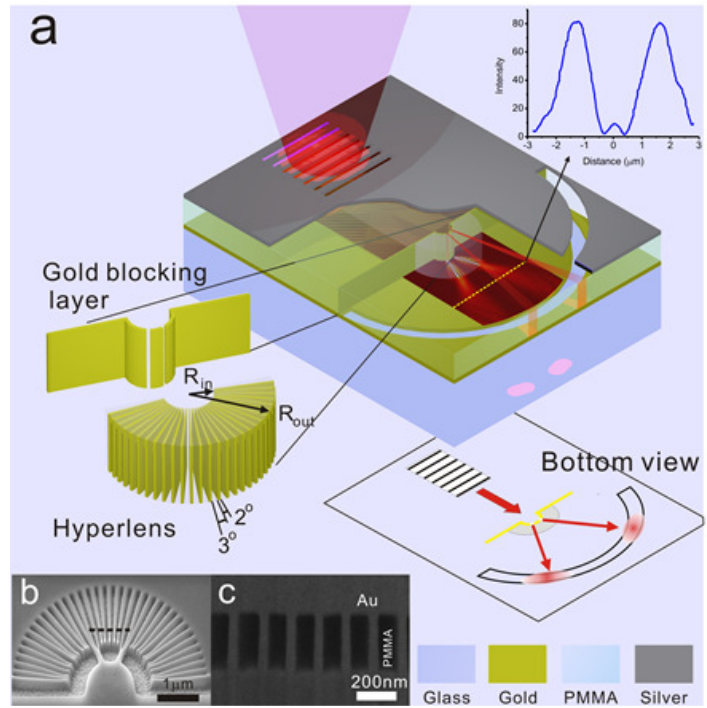

**Figure 3**

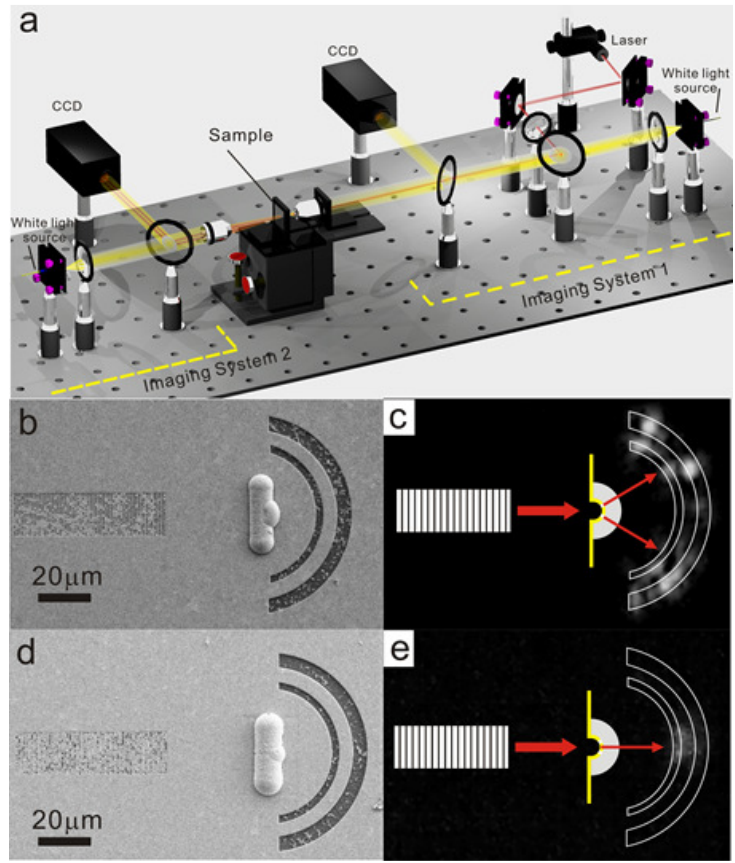

**Figure 4**

**Table**

**Table 1**.

| Mechanism | Orientation | Au/dielectric | Ag/dielectric | TiN/dielectric |
|---|---|---|---|---|
| Non-resonant | $\varepsilon_{r,\parallel}<0, \varepsilon_{r,\perp}>0$ | Visible | Visible | $\lambda>640$nm |
| Resonant | $\varepsilon_{r,\parallel}>0, \varepsilon_{r,\perp}<0$ | N/A | UV only | 510 ~ 640nm |

**Supplementary Materials:**

Materials and Methods

Figures S1-S5

References (S1-S4)

**Materials and Methods:** Figure S1 shows the schematic of the waveguide-coupled radial hyperlens and the main steps of the sample fabrication process (1)-(5). The insets show the hyperlens and the blocking layer with two nano-slits that serve as the 'object' to be imaged. In step (1), a 400nm thick Poly(methyl methacrylate) (PMMA) layer was spin-coated on top of the 300nm thick gold layer on a glass substrate. Next, in step (2), fan-shaped walls and an inner hyperlens blocking layer were patterned in the PMMA layer using the standard Electron Beam Lithography (EBL) technique (**Vistec EBPG 5000+**). Subsequently, PMMA between the walls was removed during the development and lift-off procedures. After the development, the bottom Au layer was exposed in the areas where the PMMA was removed (empty spaces between the fan-shaped PMMA pattern and the blocking layer). These empty parts were then filled with gold using the electroplating method (CHI 660A Electrochemical workstation and Techni-gold 25 es solution, Technic, Inc.) in step (3). The gold layer under the PMMA was connected to a cathode, while a Pt net was used as the anode. Since the PMMA in the patterned area was removed, the gold layer there was exposed to the solution. After the electroplating, the gaps between the PMMA walls were completely filled with gold. In step (4), a 300 nm thick silver layer was deposited on the top of the whole structure in order to create the upper metal layer of the metal-insulator-metal (MIM) waveguide. Finally, a grating coupler and two half-circular slots were made in step (5) using a focused ion beam (FIB) system (Zeiss AURIGA CrossBeam Workstation) to create the input and output ports, respectively.

The same steps were repeated for the fabrication of the reference sample that comprised exactly the same structure, waveguide, in- and out-coupling ports, and blocking layer with the slits, but no radial hyperlens. SEM images of the samples are shown in Fig. S2. Figures S2 a and b show the PMMA structures of the reference sample and the hyperlens sample patterned using the EBL step, respectively. During the electroplating, gold completely filled empty spaces between the PMMA walls and in the space where blocking layers had been pre-patterned; then, excess gold formed a capsule-shaped ridge. We used the FIB to cut the samples along the dashed lines in Fig. S2 a and d. The cross-sections of the structures are shown in Fig. S2 c and f. These cross-sections confirm that there was no layered structure observed in the reference sample as shown in Fig. S2 c, while vertical layered structure is clearly visible in the hyperlens structure as shown in Fig. S2 f.

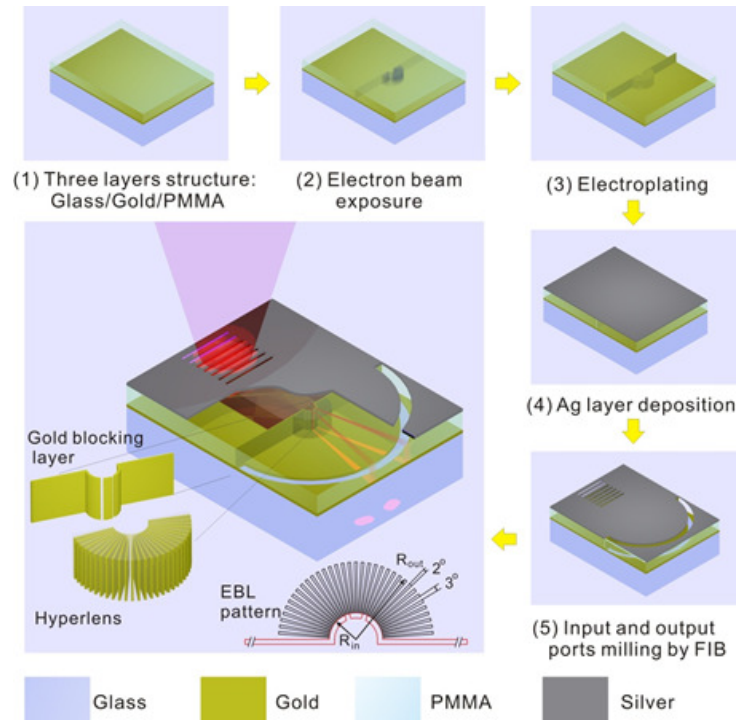

Figure S1. Center: schematic of waveguide-integrated radial hyperlens. The insets show the hyperlens and the blocking layer with two nano-slits that serve as the 'object' to be imaged. The fabrication steps are shown in side-figures (1) through (5).

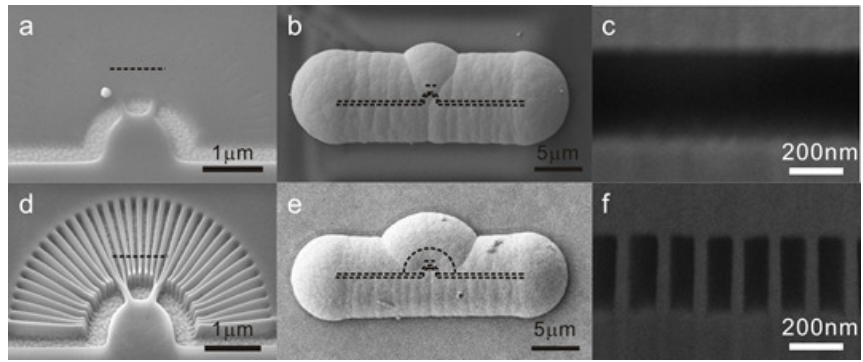

Figure S2. a and d show top views of the reference and hyperlens samples after the EBL step, respectively. b and e show the structures after the electroplating step, during which, gold completely filled empty spaces between the PMMA walls and in the space where blocking layers had been pre-patterned; then, excess gold formed a capsule-shaped ridge. c and f show the SEM images of the cross-sections of the reference sample and hyperlens sample, respectively.

**Material properties and numerical simulations:** Figure S3 shows the imaginary part of dielectric permittivity components along the layers and perpendicular to the layers of the multilayered structure calculated using the Maxwell-Garnett theory for the three cases: Au/dielectric, Ag/dielectric, and the TiN/dielectric. Permittivity data of gold and silver are taken from Ref. (S1), and TiN permittivity is from Ref. (S2). Dashed black curves show $\lambda_{pe}(f)$ for $\varepsilon_{//}$ in these three cases. Regions enclosed by the dotted black curves are the ranges of negative $\varepsilon_{\perp}$ corresponding to the resonances, associated with high resonant losses, as shown in Fig. S3 b, d, and f. For TiN, resonant loss is much larger than that for Au and Ag since its materials loss is higher in the visible wavelength range. The imaginary parts of $\varepsilon_{//}$ in these three cases are also dominated by the properties of metallic constituents of the multilayers. Figure S3a shows that $\varepsilon_{//}$ of Au/dielectric multilayer has a relatively large imaginary part around 400nm, which is due to the interband transition of gold in that wavelength range[S3]. In Fig. 3e, the imaginary part of $\varepsilon_{//}$ is very large in the whole range, caused by relatively high materials loss of TiN.

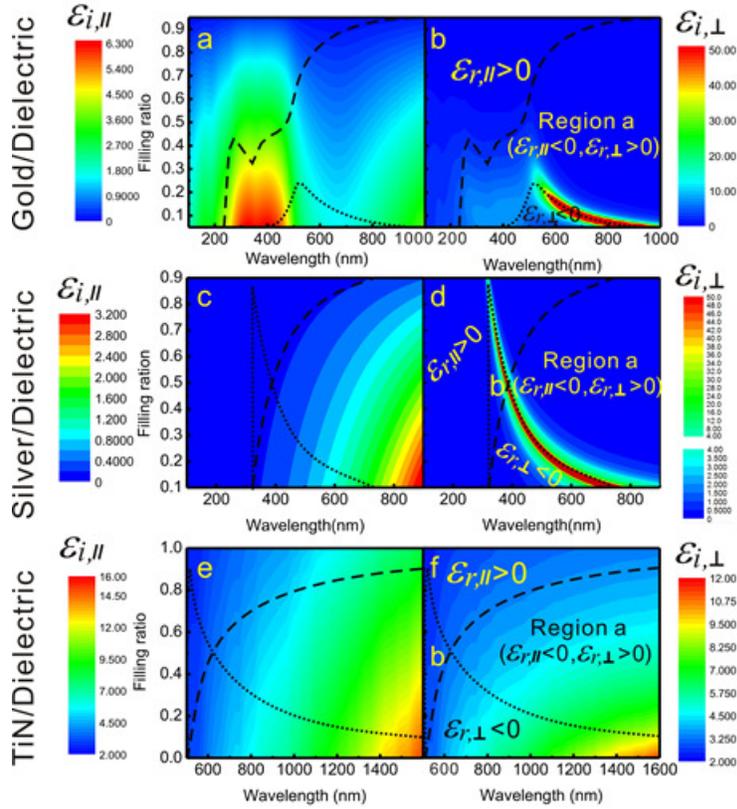

Figure S3 Imaginary part of parallel ($\varepsilon_{r,\parallel}$) and perpendicular ($\varepsilon_{r,\perp}$) components of effective dielectric permittivity as a function of filling ratio (*f*) and wavelength (λ) for three materials systems: gold/dielectric (top row), silver/dielectric (middle row), and TiN/dielectric (bottom row). The dashed black curves indicate the $\lambda_{pe}(f)$ for $\varepsilon_{r,\parallel}$ in these three cases. The negative real part of $\varepsilon_{r,\perp}$ regions resulting from the Lorentz resonance are enclosed by the dotted black curves.

In the long-wavelength limit, when the feature size of the structure is significantly smaller than the wavelength of incident light, the metamaterial with Au/PMMA alternating layers can be described as an effective uniaxial material. In our design, the radial hyperlens contains 35 pairs of Au and PMMA. Each pair occupies approximately a 5° section. In each pair, Au constitutes about 2°, and the PMMA takes up about 3°. Therefore, the filling fraction of the PMMA is 60%. Using the Maxwell-Garnett theory, we calculated the components of dielectric permittivity along the layers and perpendicular to the layers at $\lambda$=780nm to be $\varepsilon_r = -8.93+0.87i$ and $\varepsilon_\theta = 3.65+0.0178i$, respectively. A grating coupler was devised on the silver layer[S4]. The core thickness of the MIM waveguide was designed to be 400nm, in order to enable the fundamental TE mode propagation inside the waveguide. We performed three-dimensional numerical simulations of light propagation in the entire structure consisting of the grating coupler, hyperlens, PMMA layers, and the waveguide using COMSOL Multiphysics[TM] 4.3b. The results are shown in Fig. S4.

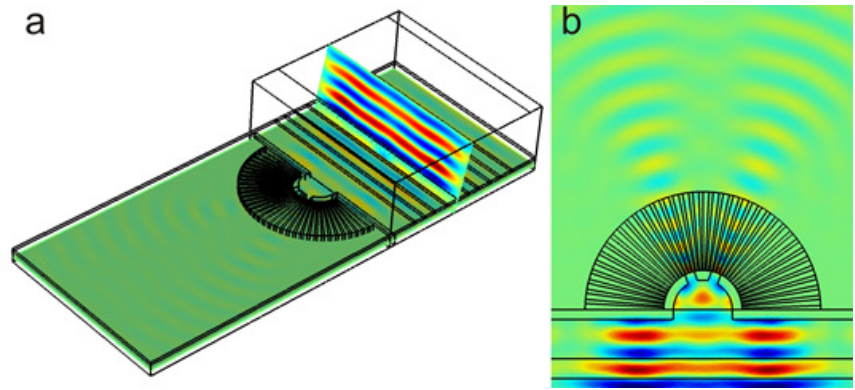

Figure S4 Simulation setup (a) and the results of light propagation in an integrated system of an MIM waveguide and a hyperlens (b).

**Optical measurement details:** The twin imaging system was employed to characterize the sample. The layout of the sample is illustrated in Fig. S5a. We utilized Imaging System 1 to locate the position of the grating coupler on the input (top) side and then coupled 780nm laser to the waveguide through that grating coupler. The output light was out-coupled through the bottom arc-shaped slits (Fig. S5a). Lower Figure S5b reveals the focused beam on the grating from the input (top) side. Upper Figure S5b is the image of the output beams on the bottom side. Note that Fig. S5b clearly shows the two output beams resolved by the hyperlens.

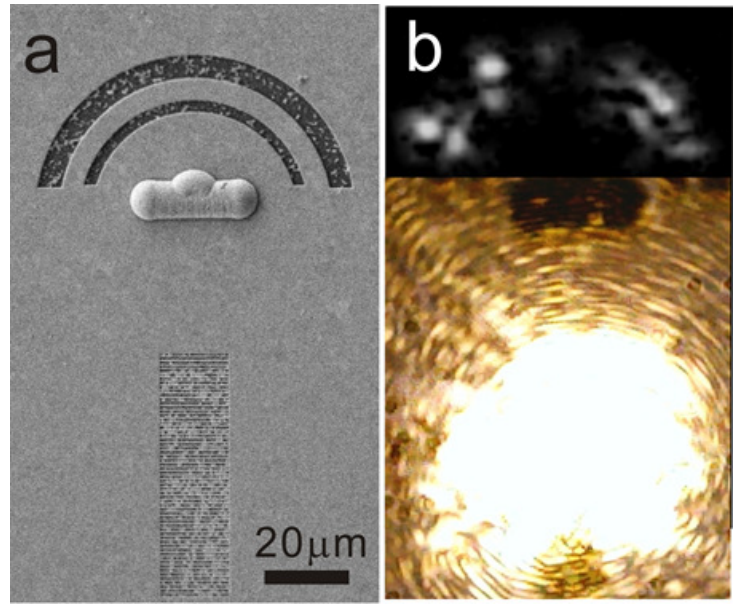

Figure S5 Layout of the sample: SEM image of the structure from the top side (a); Image of the output beams resolved by the hyperlens on the bottom side (upper image b); Image of the focused beam on the grating from the top side (lower image b).